\documentclass[twocolumn]{aastex631}

\usepackage{graphicx,times}             %for PS/EPS graphics inclusion, new
\usepackage{natbib}
\usepackage{amssymb,amsmath}
\usepackage{amsmath}
\usepackage{txfonts}
\bibpunct{(}{)}{;}{a}{}{,}
\newcommand{\ie}{{\it i.e.\ }}

\newcommand{\s}{\ensuremath{\mbox{~s}}}

\newcommand{\cm}{\ensuremath{\mbox{~cm}}}

\newcommand{\pcmcu}{\ensuremath{\cm^{-3}}}

\newcommand{\vel}{km\,s$^{-1}$}

\newcommand{\egcite}{\citep[e.g.,][]}

\begin{document}

\title{The evolution of molecular clouds: turbulence-regulated global radial  collapse}

\shorttitle{global radial  collapse}
\shortauthors{An-Xu Luo et.al}
% \pagerange{\pageref{firstpage}--\pageref{lastpage}} 
% \pubyear{2023}

% \maketitle

% \correspondingauthor{An-Xu Luo}
% \email{luoax@bao.ac.cn}
\correspondingauthor{Hong-Li Liu}
\email{hongliliu2012@gmail.com}
\correspondingauthor{Jin-Zeng Li}
\email{ljz@nao.cas.cn}

\author{An-Xu Luo}
\affiliation{School of physics and astronomy, Yunnan University, Kunming, 650091, PR China \\}
\affiliation{National Astronomical Observatories, Chinese Academy of Sciences, Beijing, 100101, China \\}
\affiliation{Both authors contributed equally to this work.\\}

\author{Hong-Li Liu}
\affiliation{School of physics and astronomy, Yunnan University, Kunming, 650091, PR China \\}
\affiliation{Both authors contributed equally to this work.\\}

\author{Jin-Zeng Li}
\affiliation{National Astronomical Observatories, Chinese Academy of Sciences, Beijing, 100101, China \\}

\begin{abstract}
The star formation efficiency (SFE) measures the proportion of molecular gas converted into stars, while the star formation rate (SFR) indicates the rate at which gas is transformed into stars. Here we propose such a model in the framework of a turbulence-regulated global radial collapse in molecular clouds being in quasi-virial equilibrium, where the collapse velocity depends on the density profile and the initial mass-to-radius ratio of molecular clouds, with the collapse velocity accelerating during the collapse process. This simplified analytical model allows us to estimate a lifetime of giant molecular clouds of approximately $0.44-7.36 \times 10^7\, \rm{yr}$, and a star formation timescale of approximately $0.5-5.88 \times 10^6\, \rm{yr}$. Additionally, we can predict an SFE of approximately $1.59\, \%$, and an SFR of roughly $1.85\, \rm{M_{\odot} \, yr^{-1}}$ for the Milky Way in agreement with observations.
\end{abstract}


\section{Introduction} \label{sec:intro}
For several decades, the birth of stars within the Milky Way's molecular clouds (MCs) has remained at the forefront of astrophysical inquiry \egcite{Mck07, Vaz19, Lu22, Li24}. Central to this research are the enigmas surrounding the low star formation efficiency (SFE) of approximately $1\%$ and the low star formation rate (SFR) of $ 1-3 \,\rm{M_{\odot} \, yr^{-1}}$ in the Milky Way \egcite{Shu87, Mck07, Cho11, Dav11, Ken12, Lic15, Eli22}. The pursuit of the corresponding answers has led to the proposition of theoretical frameworks that shed light on the mechanisms dictating the transformation of MCs into stars. 

Various theoretical models and simulations have made substantial efforts to explain the low SFE and SFR \egcite{Kru05, Pad11, Vaz19, Pad20}. For instance, the global hierarchical collapse (GHC) model emphasizes that gravity dominates the evolution and dynamics of MCs \citep{Zam14, Vaz19}. This model suggests that the SFR increases over time, leading star-forming clouds to evolve toward higher SFRs and produce stars with greater average masses. Eventually, feedback from the massive stars begins to disrupt the clouds. Regions that are far enough away, where the infall-motion front has not yet arrived, become decoupled from the collapse due to this feedback. Consequently, this material does not contribute to the star formation episode in the region, keeping the overall SFR low, as observed.
In contrast, the ``inertial inflow" model highlights the role of supersonic turbulence in massive star formation \citep{Pad20}, considering the turbulence as the major factor for both low observed SFE and SFR. However, neither model offers a simple, analytical solution to the low SFE and SFR observed in the Milky Way. Furthermore, gravoturbulent (GT) models can explain the low SFR, but involve numerous parameters \citep{Kru05}, meaning a direct comparison with observed SFE and SFR not always easy.

In this context, we present a new, streamlined semianalytical model, termed the turbulence-regulated global radial collapse (TRGRC) model. It hypothesizes that an MC, characterized by spherical symmetry in its density distribution, is subject to a global collapse within an isotropic and supersonically turbulent environment. Additionally, the model assumes (1) the MC exists in (near) isolation and unaffected by magnetic fields, (2) it retains a quasi-static state, upholding a quasi-virial equilibrium, and (3) the supersonic turbulence is isotropic. Note that real MCs are not isolated, and their turbulence at small scales (e.g., $\sim 0.01$\,pc) deviates from isotropy. Nevertheless, these intricacies are overlooked to maintain the model's simplicity. Within this model framework, a spherical MC is envisaged to undergo a global, quasi-static collapse radially, driven by its self-gravity and regulated by non-negligible turbulence. That is, the supersonic turbulence introduces random local motions that hinder but not halt the MC's global radial collapse. The initial velocity at the onset of collapse strongly depends on the MC's density distribution and its initial mass $M_{0}$ and radius $R_{0}$.

\section{The global radial collapse}

\begin{figure*}[ht!]
\centering
\includegraphics[width=0.7\linewidth]{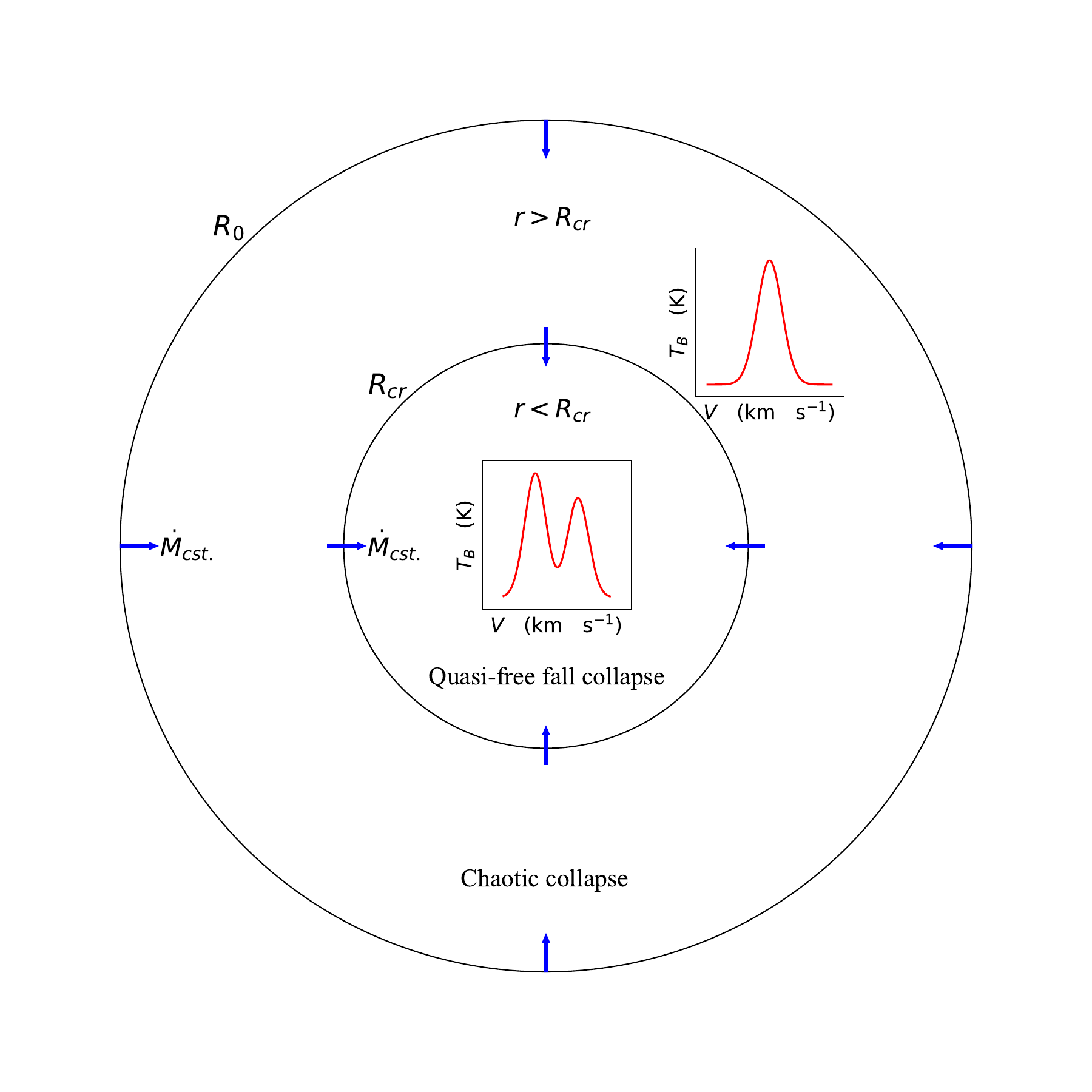}
\caption{An illustration of the gas radial collapse. The $R_{0}$ represents the boundary of MC, $R_{cr}$ represents the critical transition scale, and $\Dot{M}_{cst.}$ represents the constant mass infall rate. In regions where $r > R_{cr}$, the gas undergoing chaotic collapse and the Gaussian spectral lines are readily observed. In regions where $r < R_{cr}$, the gas undergoing quasi-freefall collapse and the blue asymmetric spectral lines are easily observed.}
\label{fig:model plot}
\end{figure*}

\subsection{Constant mass infall rate $\Dot{M}_{cst.}$}
We consider MCs as supersonically turbulent gas flows governed by the mass continuity equation:
\begin{align}
    \frac{\partial \rho}{\partial t} + \nabla \cdot (\rho \Vec{u}) = 0,
\end{align}
where $\partial \rho / \partial t$ represents the rate of change of density with time, and $\Vec{u}$ is the flow velocity.
As gas from an outer shell accretes onto an inner shell, gas from an even larger shell simultaneously moves inward to replenish the outer shell. This continuous inward flow maintains a roughly constant gas density at any given radius, implying a state of quasi-static process. While the central density theoretically increases toward infinity, we neglect the physical implications of this extreme density for the sake of simplicity. The quasi-static approximation of MCs implies $ \partial \rho / \partial t = 0$, leading to $\nabla \cdot (\rho \Vec{u}) = 0$ with $\Vec{u} = (u_r, u_{\theta}, u_{\varphi})$ in a spherical coordinate system. Assuming a spherically symmetric density distribution allows us to consider collapse occurring only in the radial direction, yielding $\Vec{u} = u_r$. Therefore,
\begin{align}\label{eq:mass-cont}
    \nabla \cdot (\rho \Vec{u}) = \nabla \cdot (\rho u_{r}) = \frac{1}{r^{2}} \frac{\partial}{\partial r} (r^{2} \rho u_{r}) = 0.
\end{align}
This equation results in a constant mass infall rate 
\begin{align}\label{eq:dotM_cst}
\Dot{M}_{cst.} = r^{2} \rho u_{r}. 
\end{align}
% $\Dot{M}_{cst.} = r^{2} \rho u_{r}$. 
Consequently, the mass of gas passing through any hierarchical, spherical surface remains constant per unit time, as illustrated in Fig.\,\ref{fig:model plot}. Note that the condition $\partial \rho / \partial t = 0$ implies a net mass flux of $\dot{M}_{\rm cst.} = 0$ across a fixed radius, where gas flows both inward and outward. However, this does not imply that $\dot{M}_{\rm cst.} = 0$ for gas flowing unidirectionally toward the cloud center.

If the density distribution of spherically symmetric MCs follows a power-law profile of $\rho = \rho_{re} (r/r_{re})^{-p}$ with $\rho_{re}$ known at the reference $r_{re}$, we can derive
\begin{align}
     \Dot{M}_{cst.} = r_{re}^2 \rho_{re} u_r \left(\frac{r}{r_{re}}\right)^{2-p}.\label{Eq mass infall rate}
\end{align}
Furthermore, the radial collapse velocity $u_r$ can be expressed as
\begin{align}
     u_r = \frac{\Dot{M}_{cst.}}{\rho_{re} r_{re}^2} \left(\frac{r}{r_{re}}\right)^{p-2}. \label{collapse velocity Eq}
\end{align}
For $p = 2$, a constant velocity emerges, indicating a scale-free gravitational collapse \citep{Li18}, akin to the collapse of the envelope of a hydrostatic singular isothermal sphere \citep{Shu77}. However, when $p < 2$, $u_r$ increases as $r$ decreases, suggesting that the collapse velocity accelerates down to smaller scales where star formation occurs. This increasing trend of collapse velocity with decreasing scale has been supported by observations toward high-mass star formation regions \citep{Yan23}. Note that when $p > 2$, $u_r$ decreases as $r$ decreases. We do not consider the case of $p>2$, as the global density profile across almost the whole cloud scales has been observed to be shallower with $p<2$ \egcite{Lar81, Luo24, Zha24}.

\subsection{Critical transition scale $R_{cr}$}\label{Sect.2.2}
The density distribution $\rho = \rho_{re} (r/r_{re})^{-p}$  yields the gravitational potential energy $U$ as follows \citep{Ber92}:
\begin{align}
    U = -\frac{3-p}{5-2p} \frac{GM^2}{r}. \label{U Eq}
\end{align}
The kinetic energy $E_k$ includes contributions from supersonic turbulence ($E_{turb}$), thermal motions ($E_{ther}$), and the kinetic motions of infall gas resulting from self-gravitational collapse ($E_{a}$). Since both thermal and turbulent motions counteract the self-gravity of MCs, and with supersonic  turbulent motions significantly stronger than thermal motions ($E_{turb} \gg E_{ther}$), we combine them as $E_{t} = E_{turb} + E_{ther} \approx E_{turb}$.

 For an MC in a state of virial equilibrium, the potential energy is equal to twice the kinetic energy \citep{Ber92},
\begin{align}
    2E_k + U = 0 \Leftrightarrow \mid -U \mid = 2(E_{t} + E_{a}).\label{virial Eq}
\end{align}
Mathematically, Eq.\,\ref{virial Eq} suggests when $E_a \not = 0$, $2E_t < \mid -U \mid$ corresponding to the condition of gravitational collapse. Observationally, \cite{Luo24} found a nearly flat virial parameter--mass ($\alpha_{vir}-M$) distribution across a wide range of density scales over 0.01--100\,pc, with the virial parameter centered around  unity, suggesting a state of global quasi-virial equilibrium across multiple scales. Based on this observed fact, we hypothesize that MCs could maintain (quasi-) virial equilibrium throughout their evolution. Additionally, the observed slight dominance of gravitational pressure over turbulent pressure ($P_{\rm turb} \propto P_{\rm gra}^{0.85}$) suggests that MCs may undergo a slow, quasi-static collapse \citep{Luo24}. Therefore, we assume that a slow collapse can occur within a virial or quasi-virial equilibrium state. In the case where all gravitational energy is converted to infall kinetic energy, corresponding to $E_t = 0$ without any  support against gravity, the equation simplifies to $\mid -U \mid = 2E_{a}$. Substituting this relation into Eq.\,\ref{U Eq}, we derive the initial collapse velocity $u_{0}$ of the MCs at $R_{0}$ as follows:
\begin{align}
    u_{0} = \frac{2E_{a,0}}{M_0} = \left(\frac{3-p}{5-2p} \frac{GM_{0}}{R_{0}}\right)^{1/2}. \label{initial velocity Eq}
\end{align}

According to Eq.\,\ref{Eq mass infall rate}, when $p=2$ and the collapse is unhindered, the radial collapse velocity $u_r = u_0$ and the constant mass infall rate $\dot{M}_{\rm cst.} = r_{\rm re}^2 \rho_{\rm re} u_0$. For $p \neq 2$, $\dot{M}_{\rm cst.}$ maintains a similar form but with a different constant value dependent on the specific $\rho_{\rm re}$ and $u_0$, which vary with the index $p$. Specifically, since the gas collapse is solely driven by self-gravity, the support factors provided by thermal or turbulent motions does not induce a net mass inflow. Consequently, $\dot{M}_{\rm cst.}$ is solely determined by the initial mass $M_0$, initial radius $R_0$, and density profile $\rho = \rho_{\rm re} (r/r_{\rm re})^{-p}$ of the cloud. Eqns.\,\ref{U Eq} and \ref{initial velocity Eq} reveal that different density profiles correspond to varying gravitational potential energies and initial velocities $u_0$, which, in turn, affect the value of $\dot{M}_{\rm cst.}$. Once the density profile of a cloud is established, the mass infall rate can still be expressed in the same form as derived for the $p=2$ case. In this context, Eq.\,\ref{collapse velocity Eq} can be re-expressed as 
\begin{align}
    u_r = u_{0} \left(\frac{r}{r_{re}}\right)^{p-2}. \label{collapse velocity Eq2}
\end{align}
That is, for a given density profile $\rho = \rho_{re} (r/r_{re})^{-p}$, support factors  can slow down the speed of gas infall, but do not change the mass infall rate.

From Eq.\,\ref{collapse velocity Eq2}, the collapse velocity $u_r$ is small on large scales, making the $E_t$ become the primary contributor to the kinetic energy $E_k$. Due to the dissipation of turbulence and the acceleration of collapse velocity, $u_r$ will exceed the turbulent velocity at a critical transition scale $R_{cr}$, thereby dominating the energy balance. Consequently, we expect $R_{cr}$ to appear at the scale where the turbulent velocity equals $u_r$. If turbulent velocity follows $\sigma = \sigma_{re} (r/r_{re})^{\beta}$, where $\beta = 1-p/2$ for an MC maintaining quasi-virial equilibrium \citep{Mck03, Luo24}, the equality of the turbulent and collapse velocities gives rise to
\begin{align}
    \left. \sigma_{re} \left(\frac{r}{r_{re}}\right)^{1-p/2} \right |_{r=R_{cr}} = \left. u_{0} \left(\frac{r}{r_{re}}\right)^{p-2} \right |_{r=R_{cr}}. \label{eq10}
\end{align}
Replacing $r$ with $R_{cr}$ in this equation, we obtain
\begin{align}
    \frac{R_{cr}}{r_{re}} = \left(\frac{\sigma_{re}}{u_{0}}\right)^{2/(3p-6)}. \label{p-beta Lc Eq}
\end{align}

Additionally,  we point out that when thermal motions significantly stronger than turbulent motions ($E_{ther} \gg E_{turb}$) in subsonic regions, $E_{t} = E_{turb} + E_{ther} \approx E_{ther}$. Similarly, we expect $R_{cr}$ to appear at the scale where the thermal motion velocity $c_s$ equals $u_r$. Then, Equations\,\ref{eq10} and \,\ref{p-beta Lc Eq} can be written as
\begin{align}
    c_s = \left. u_{0} \left(\frac{r}{r_{re}}\right)^{p-2} \right |_{r=R_{cr}},
\end{align}
and
\begin{align}
    \frac{R_{cr}}{r_{re}} = \left(\frac{c_s}{u_{0}}\right)^{1/(p-2)}. 
\end{align}

For $r > R_{cr}$, where $\sigma > u_r$, turbulence dominates the velocity field, leading to chaotic collapse behavior. In this region, the gas is not expected to directly participate in the star formation process due to the randomness of the collapse. Consequently, the chaotic collapse velocity will hardly produce the blue asymmetric spectra in line emission of gravitational collapse tracers, making it difficult to observe. For $r < R_{cr}$, where $u_r > \sigma$, the collapse velocity dominates over the turbulent velocity. Thus, the collapse within $R_{cr}$ is anticipated to resemble a quasi-freefall collapse. The gas within this region can directly participate in the star formation process. If real clouds possess an isotropic and spherically symmetric density distribution, the velocity field dominated by ordered collapse motions will result in observable blue asymmetric molecular emission lines (as shown in Fig.\,\ref{fig:model plot}). This aligns with recent observational results from \cite{Xu23}, who observed a gradual transition in molecular line profiles from Gaussian at the edges of molecular clumps to blue asymmetric at the center. Note that Eq.\,\ref{p-beta Lc Eq} becomes invalid when $p = 2$, indicating the absence of a critical transition scale. This may mean that the gas throughout the entire spherical framework can participate in the formation of stars for the scenarios of scale-free gravitational collapse \citep{Li18}.

\subsection{SFE and SFR}\label{sec2.3}
Based on the collapse velocity, we can define the gas infall time as
\begin{align}
   dt =  \frac{dr}{u_r} \Leftrightarrow dt = \frac{(r/{r_{re}})^{2-p}}{u_0}dr.\label{dL/dt Eq}
\end{align}
We define the time when the gas to travel from the boundary $R_0$ to the gravitational center of MCs as $t_{MC}$:
\begin{align}
    t_{MC} = \frac{r_{re}}{(3-p)u_0} \left(\frac{R_{0}}{r_{re}}\right)^{3-p}.
\end{align}
If the collapse and evolution of an MC are (nearly) isolated, $t_{MC} $ can be taken to approximate the cloud lifetime.

We define a critical transition scale, $R_{cr}$, within which the gas is directly involved in the star formation process. Beyond this scale, while external gas can influence density distribution, and potentially contribute to star formation, turbulent motions dominate over infall, limiting the effectiveness of gas accretion. As a result, gas outside $R_{cr}$  is less likely to participate in star formation compared to that within the critical scale.
 Similarly, we express the time required for the gas from $R_{cr}$ to the gravitational center of MCs as $t_{cr}$:
\begin{align}
    t_{cr} = \frac{r_{re}}{(3-p)u_0} \left(\frac{R_{cr}}{r_{re}}\right)^{3-p}.
\end{align}
Since only the gas within the region of $r < R_{cr}$ can effectively participate in star formation, $t_{cr}$ can serve as a reasonable approximation of the star formation timescale.

Therefore, in an MC, the total gas mass available for star formation is $M_{cr} = \Dot{M}_{cst.} \cdot t_{cr}$. This mass ultimately transfers to the precursors that form stars, \ie, prestellar cores. However, due to the conversion efficiency of $\epsilon \sim 0.25-0.75$ from prestellar cores to newborn stars \egcite{Maz00, Tan17, Sta19}, the mass $M_{cr}$ relates to the stellar mass via $M_* = M_{cr} \cdot \epsilon$. The SFE over the time interval $t_{MC}$ can then be determined as
\begin{align}
    \mathrm{SFE} = \frac{M_*}{M_0}. \label{SFE Eq}
\end{align}
Given $M_* = M_{cr} \cdot \epsilon = \Dot{M}_{cst.} \cdot t_{cr} \cdot \epsilon$, Eq.\,\ref{SFE Eq} indicates that an extended $t_{cr}$ (depending on the density profile, $\sigma_{re}$, and $u_0$) and an increased $\Dot{M}_{cst.}$ (depending on the density profile and $u_0$) enhance the efficiency with which an MC  converts its gas into new stars.

Moreover, the SFR for an MC over their lifetimes can be expressed as
\begin{align}
    \mathrm{SFR} = \frac{M_*}{t_{MC}} = \mathrm{SFE} \cdot \frac{M_0}{t_{MC}}. \label{SFR Eq}
\end{align}
Recalling \( M_* = \Dot{M}_{cst.} \cdot t_{cr} \cdot \epsilon \),  Eq.\,\ref{SFR Eq} suggests an increase in \( t_{cr} \) or \( \Dot{M}_{cst.} \) serves to accelerate the star formation process. This depends on the enhanced self-gravity of MCs, leading to a faster collapse and an earlier initiation of gas into star formation. As a result, both a high SFR and SFE are the natural outcomes of the strong self-gravity of MCs, which is further influenced by the clouds' density structure and the level of turbulence.

Assuming that MCs in the Milky Way have similar SFEs, we can estimate the SFR of the Milky Way over the period \( t_{MC} \) as:
\begin{align}\label{Milk SFE Eq}
    \mathrm{SFR_{MW}} \simeq \mathrm{SFE} \cdot \frac{M_\mathrm{{gas}}}{t_{MC}},
\end{align}
where $M_\mathrm{{gas}}$ is the total mass of MCs in the Milky Way.

\section{Comparison with observations}\label{sec3}
Our model can be used to elucidate various observable parameters. For illustration, we leverage data from 158 giant molecular clouds (GMCs) observed in the Boston University-FCRAO Galactic Ring Survey (GRS) \egcite{Hey09}. The dataset reflects a global quasi-virial equilibrium state for GMCs and unveils a density profile for them of $\rho = \rho_{re} (r/r_{re})^{-1.54}$, alongside a turbulent velocity profile of $\sigma = \sigma_{re} (r/r_{re})^{0.26}$, where $\rho_{re}$ is $2.2 \times 10^{-20}$ \,g\,\pcmcu, $\sigma_{re}$ is $1.16$\,\vel, and $r_{re}$ is 1 \,pc \citep{Luo24}. By integrating these empirical relationships into our framework, we have computed several key parameters for the GMCs, which are summarized in Table\,\ref{tab: table1} and compiled in Appendix Table\,\ref{Table}.

\begin{center}
\begin{table}[]
    \centering
    \caption{Summary of the main parameters for 158 GMCs.}
    \begin{tabular}{c|c|c|c|c|c}
    % \hline
    \hline
        Parameter  &       Unit            &  Mean  &  Median  &  Max   & Min   \\
        \hline
        $\Dot{M}_{cst.}$ & $10^{-3}\rm{\,M_\odot\,yr^{-1}}$ & 1.04 & 1.0 & 3.08   & 0.33 \\
        % \hline
        $R_{cr}$      &              pc         & 4.52  &  3.86   &  20.35  & 0.81  \\
        $t_{MC}$   &  $10^{7}\,\rm{yr}$  & 1.71  &  1.73   &  7.36  & 0.44  \\
        $t_{cr}$      &  $10^{6}\,\rm{yr}$  & 1.56  &  1.63   &  5.88  & 0.50  \\
        $M_{*}$    &     $\rm{M_\odot}$  & 1166   &  800    & 9061   & 82  \\
        SFE        &        $\%$            & 1.86   & 1.59    & 6.01   & 0.36 \\ 
        SFR        & $\rm{10^{-5}\,M_\odot\,yr^{-1}}$ & 6.96 & 4.28 & 67.62   & 0.42 \\
         \hline
    \end{tabular}
    \label{tab: table1}
\end{table}
\end{center}

The transition of MCs from a state of chaotic collapse to a quasi-freefall collapse occurs at a critical radius, $R_{cr}$. Analysis of 158 GMCs reveals that $R_{cr}$ has a median of $3.86$\,pc and an average of $4.52$\,pc, covering a range between 1 and 6\,pc (refer to Appendix Fig.\,\ref{fig:parameter distribution plot} b) which corresponds to the clump scales \egcite{Hey15, Bal20, Per23}. The time $t_{MC}$, varies from $0.44-7.36 \times 10^7 \,\rm{yr}$, and the critical time, $t_{cr}$, spans $0.5-5.88 \times 10^6 \,\rm{yr}$, both displaying similar distributions. Note that the average and median $t_{MC}$ are $1.71 \times 10^7\,\rm{yr}$ and $1.73 \times 10^7 \,\rm{yr}$, respectively, while for $t_{cr}$, they are $1.56 \times 10^6\,\rm{yr}$ and $1.63 \times 10^6 \,\rm{yr}$, respectively. Therefore, we anticipate that the lifetime of GMCs could be approximately $10^7\,\rm{yr}$, as confirmed by previous studies \egcite{Che20, Jef21, Che23}. We also expect the star formation timescale to be around $10^6\,\rm{yr}$, which aligns with simulation results \egcite{Pad20, Gru22}. Nevertheless, further observations are needed to validate these predictions.

% \subsection*{SFE and SFR in the Milk Way} 
Our model assumes a constant mass infall rate (see Eq.\,\ref{eq:dotM_cst}), which is approximately $10^{-3}$ $\rm{M_\odot\, yr^{-1}}$ for the GMCs in our Milky Way (see Table\,\ref{tab: table1}). Naturally, GMCs with different initial conditions exhibit varying mass infall rates, with differences reaching an order of magnitude ($\Dot{M}_{cst.} \sim 0.33-3.08 \times 10^{-3}\, \rm{M_\odot\, yr^{-1}}$). Based on Eq.\,\ref{SFE Eq} and assuming a typical value of $\epsilon \simeq 0.5$, we estimated the SFEs for 158 GMCs in our Milky Way. The mean and median SFEs are 1.86\,\% and 1.59\,\%, respectively, consistent with the low SFE observed for the Milky Way \egcite{Kru07, Mck07}. Furthermore, the SFRs for 158 GMCs vary from $0.42-67.62 \times 10^{-5} \,\rm{M_\odot\, yr^{-1}}$, which are in agreement with observations \citep{Lad10}.

Likewise, assuming a uniform spatial distribution of GMCs in the Milky Way, we can estimate the average SFR for the Milky Way using Eq.\,\ref{Milk SFE Eq} as follows:
\begin{align}
    \mathrm{SFR_{MW}} \simeq 0.0159 \cdot \frac{2 \times 10^9 \, \rm{M_{\odot}}}{1.73 \times 10^{7} \,\rm{yr}} = 1.85 \,\rm{M_{\odot}\, yr^{-1}}.
\end{align}
Here, we utilized the median values of SFE and $t_{MC}$ for the 158 GMCs, along with its total molecular gas mass of $2 \times 10^9 \, \rm{M_{\odot}}$ \egcite{Fer01, Kle10}. The resulting average SFR agrees with current observational results \egcite{ Cho11, Dav11, Lic15, Eli22}.

\begin{figure}[t!]
\centering
\includegraphics[width=3.3 in]{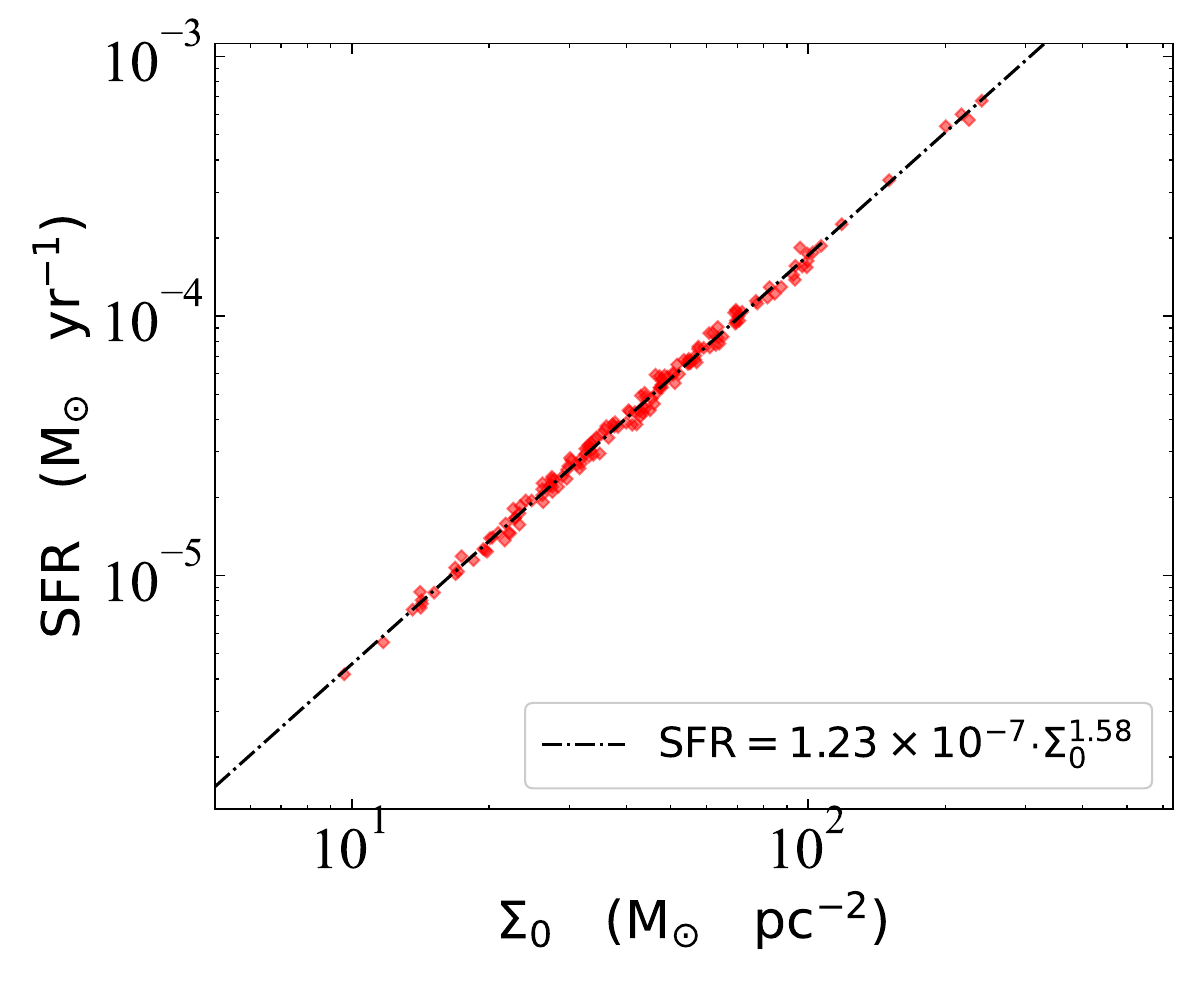} 
\caption{The relation between SFR and gas surface density for the 158 GMCs. The dashed-dotted line is the linear regression fit to the data points.}
\label{fig:K-S re plot}
\end{figure}

Moreover, our model allows us to predict the relationship between the SFR and the initial gas surface density for GMCs:
\begin{align}\label{K-S Eq1}
    \mathrm{SFR} \propto \Sigma_{0}^{\frac{12-5p}{12-6p}} \cdot R_{0}^\frac{-24+25p-6p^2}{12-6p},
\end{align}
where the initial gas surface density is $\Sigma_{0} = M_0/(\pi R_{0}^2)$. For $p = -1.54$, the power index $\frac{12-5p}{12-6p} \simeq 1.56$. However, when $p = 1.5$, $\frac{-24+25p-6p^2}{12-6p} = 0$, indicating that the SFR depends solely on $\Sigma_{0}$. Given a conversion efficiency $\epsilon = 0.5$, the relationship between SFR and $\Sigma_{0}$ for the GMC sample from the GRS survey is (see Fig.\,\ref{fig:K-S re plot}):
\begin{align}\label{K-S GNC}
    \mathrm{SFR} \simeq 1.23 \times 10^{-7} \cdot \left(\frac{\Sigma_{0}}{\mathrm{M_{\odot} \cdot pc^{-2}}}\right)^{1.58} \,\rm{M_{\odot} \cdot yr^{-1}}.
\end{align}
This predicted SFR--$\Sigma_{0}$ relationship highlights that denser initial conditions of MCs lead to higher SFRs, warranting further validation by future observations. It is worth noting that a power-law relationship between the SFR per unit area of galaxies and their gas surface density, \ie, the Kennicutt-Schmidt (K-S) law \citep{Ken98}, has been widely discussed. However, our TRGRC model cannot reproduce the empirical K-S relation, which will be addressed in future work.

It should be noted that further observations are needed to validate the density profile and turbulent velocity scaling relation we used. The observed turbulent velocity scaling relation is generally of the form $\sigma \propto r^{\beta}$, with a power-law exponent $\beta \sim 0.5$, which is higher than 0.26 adopted in this work. The latter value was interpreted as a possible result of added contribution of magnetic fields to turbulence. However, different turbulent velocity scaling relations may correspond to different density distributions \egcite{Lar81, Mck03, Luo24}. Therefore, we expect that if MCs are in quasi-virial equilibrium, applying different density distributions and turbulent velocity scaling relations to the TRGRC model can still yield similar results.

\section{Comparison with other models and caveats of our model}
Our TRGRC model shares several similarities with the GHC model \citep{Vaz19}, particularly in the scenario that MCs are undergoing a global collapse. However, the GHC model assumes moderately supersonic turbulence (with a typical Mach number of about 3) and neglects its sufficient influence on MC dynamics \citep{Vaz19, Vaz24b}. Our model emphasizes the significant role of turbulence in regions where turbulence is too strong not to be ignored for example in large-scale MCs.
% In addition, our model proposes that the low SFE and low SFR could be primarily a result of the MCs self-gravity and we incorporates the influence of feedback .
Our model suggests that MCs with strong self-gravity naturally exhibit higher SFE, while those with weaker self-gravity tend to have lower SFE; both low SFE and SFR arise from the interplay between self-gravity, turbulence, and stellar feedback. In contrast, the GHC model attributes the low SFE  to stellar feedback alone \citep{Vaz19}.

In the GT model, supersonic turbulence provids support against self-gravity, which aligns with our current understanding. Additionally, the GT model proposes that MCs do not collapse on large scales, with collapse occurring only at the smallest scales (e.g., molecular cores with density  $n >10^{4}$ \pcmcu\ and size about 0.1\,pc, and even smaller scales) \citep{Pad02, Kru05}. However, our model presents a different perspective, demonstrating the global collapse of MCs undertaken across almost all scales under conditions of supersonic turbulent support.

It is worth noting that past kinematics observations toward molecular cores \egcite{Goo98, Pin10, Fri17} suggest that at a critical scale $R_{\mathrm{coh}} \sim 0.1$ pc, where the coupling between the magnetic field and gas motion significantly decreases, the gas motion transitions from supersonic turbulence to subsonic state, leading to more ordered motions. This is similar to our TRGRC model, but our model does not consider the influence of magnetic fields, which should be cautioned. \cite{Vaz24a} proposed that the mass growth of a core is regulated by the 'gravitational choking' mechanism, which is ideally driven by gravity without any additional support against it. While both the 'gravitational choking' mechanism and the TRGRC model agree that gravity regulates gas infall across all cloud scales, the TRGRC model incorporates the support against the selt-gravity of MCs from turbulence in large-scale cloud structures.

To what scenario is the TRGRC model applicable: spherical MC or filament? The diversity of geometric morphologies observed in MCs requires a more comprehensive and versatile model to describe their evolution. In our model, we did not address this geometric diversity. We developed a simplified analytical framework based on the assumption of spherical symmetry to approximately describe the global collapse of MCs. This assumption was chosen not only for its intuitive simplicity, but also it significantly simplifies the mathematical formulation and derivations. Our model is built in the context of the global radial collapse regulated by turbulence. The extension of its applicability to other-shaped MC structures by adjusting the geometric assumptions, such as cylindrical (filamentary) configurations warrants future dedicated investigations, which is beyond the scope of this paper.

To what scale is the TRGRC model applicable: large-scale cloud structures or smallest-scale clumps/cores? The TRGRC model has been applied to GMCs in quasi-virial equilibrium (see Sect.\,\ref{sec3}), yielding predicted results consistent with observations, such as both low SFE and low SFR. The TRGRC model describes a single-collapse scenario across all cloud scales, but star formation often involves the formation of multiple star systems even clusters that may experience different modes of fragmentation on different scales, such as a cylindrical fragmentation mode in filaments and a nearly spherical Jeans-like mode in smaller-scale clumps \egcite{Kai13, Kai17}. Due to the absence of suitable fragmentation mechanisms considered in the TRGRC framework to fully capture the evolution of MCs, caution is made when considering the TRGRC model to explain the fragmentation properties of MCs. If the mass infall rate remains constant during the fragmentation process, we suggest that the TRGRC model could be applicable to both MCs and clumps/cores. We predict that the evolution of MCs will undergo a cyclic process of collapse-fragmentation-collapse, and we will further explore in a forthcoming paper the fragmentation mechanisms of MCs within the TRGRC model.

\section{summary}\label{conclusions}
Our new semianalytical model describes the global radial collapse of an MC. It emphasizes that the MC in quasi-virial equilibrium is always undergoing global collapse in the radial direction, while highlighting the different collapse regions and that the variations in collapse modes are regulated by turbulence. The TRGRC model has successfully reproduced the low SFE and SFR observed within the Milky Way. Additionally, this model facilitates the prediction of various observational indices, including the lifespan of GMCs, star formation time, and both the SFE and SFR for individual  MCs. Future validation of this model with new observational data will refine the turbulence-regulated global radial collapse paradigm, providing promising insights into the evolution of molecular clouds into stars.

\vspace{5mm}
\noindent    
% \begin{acknowledgments}
We thank the anonymous referee for comments and suggestions that greatly improved the quality of this paper.  This work was supported by the Key Project of Inter national Cooperation of Ministry of Science and Technology of China through grant 2010DFA02710, and by the National Key R\&D Program of China (No.\,2022YFA1603101). H.-L. Liu is supported by National Natural Science Foundation of China (NSFC) through the grant No.\,12103045, by Yunnan Fundamental Research Project (grant No.\,202301AT070118, 202401AS070121), and by Xingdian Talent Support Plan--Youth Project. Part of this work was carried out  at the China-Argentina Cooperation Station of NAOC/CAS.
% \end{acknowledgments}

\appendix \label{appendix}

% \section{appendix A} {appendix a}

\begin{figure*}
\centering
\includegraphics[width=6.5 in]{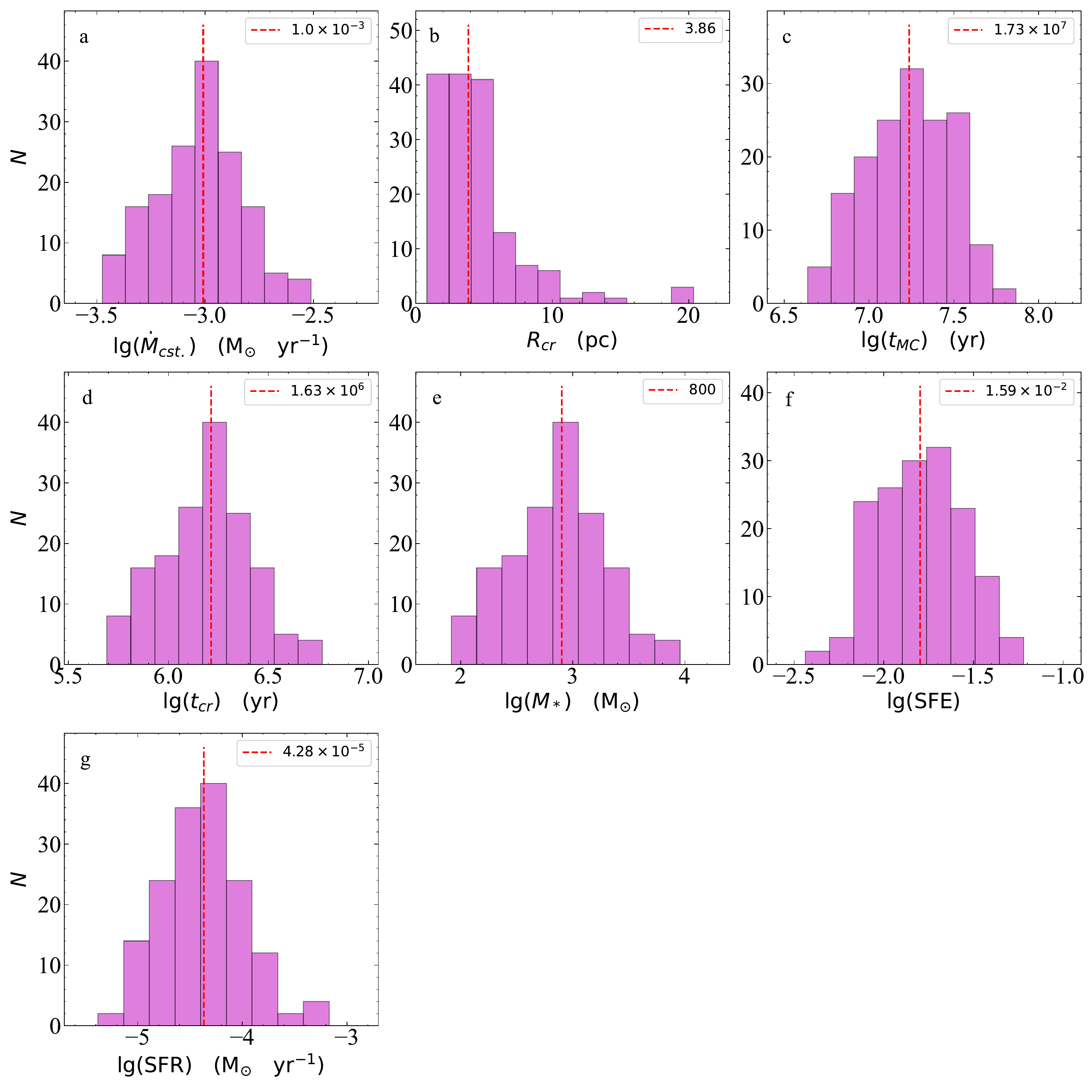} 
\caption{The histogram distribution of seven important parameters. The red vertical dashed line indicates the median for each parameter.}
\label{fig:parameter distribution plot}
\end{figure*}

\clearpage

\startlongtable
% \begin{center}\label{Table}
\begin{deluxetable*}{ccccccccccc}\label{Table}
\tabletypesize{\normalsize}
% \tablewidth{0pt} 
\tablecaption{The physical parameters of 158 GMCs} 
\tablehead{
\colhead{GMC} & \colhead{$R_0$} & \colhead{$M_0$} & \colhead{$u_0$} & \colhead{$\Dot{M}_{cst}$} & \colhead{$R_{cr}$} & \colhead{$t_{MC}$} & \colhead{$t_{cr}$} & \colhead{$M_*$}  & \colhead{SFE} & \colhead{SFR}  \\
      & \colhead{pc} & \colhead{$\rm{M_{\odot}}$}   &  \colhead{\vel} & \colhead{$\rm{M_{\odot}\, yr^{-1}}$} & \colhead{pc} &  \colhead{yr} & \colhead{yr} & \colhead{$\rm{M_{\odot}}$}  &   & \colhead{$\rm{M_{\odot}\, yr^{-1}}$}  \\
        &  & \colhead{$10^{4}$} &  & \colhead{$10^{-3}$} &   & $10^{7}$ & $10^{6}$ &  & \colhead{$10^{-2}$}   & \colhead{$10^{-5}$} }
\startdata 
  GMC 64   &  40.6   &   32.0   &  5.08   &  1.69   &    8.5   &  2.94    &   3.0    &  2531    &  0.79    &    8.6  \\ 
  GMC 67   &  42.2   &    7.9   &  2.47   &  0.82   &    3.0   &  6.39    &  1.34    &   553    &   0.7    &   0.87  \\
  GMC 68   &   6.9   &    1.4   &  2.58   &  0.86   &   3.18   &  0.44    &  1.41    &   602    &   4.3    &  13.81  \\ 
  GMC 70   &   9.7   &    1.3   &  2.09   &   0.7   &   2.35   &  0.88    &  1.12    &   388    &  2.99    &    4.4  \\ 
  GMC 71   &  10.3   &    1.9   &  2.46   &  0.82   &   2.97   &  0.82    &  1.33    &   544    &  2.87    &   6.63  \\ 
  GMC 72   &  41.3   &   26.0   &  4.54   &  1.51   &   7.22   &  3.38    &  2.65    &  1995    &  0.77    &   5.91  \\ 
  GMC 73   &   4.6   &   0.28   &  1.41   &  0.47   &   1.33   &  0.44    &  0.72    &   168    &  6.02    &   3.82  \\ 
  GMC 74   &  32.1   &   11.0   &  3.35   &  1.11   &   4.65   &  3.17    &  1.88    &  1048    &  0.95    &   3.31  \\ 
  GMC 75   &  23.2   &   13.0   &  4.28   &  1.42   &   6.64   &  1.54    &  2.48    &  1764    &  1.36    &  11.44  \\ 
  GMC 77   &  33.8   &   12.0   &  3.41   &  1.13   &   4.77   &  3.35    &  1.92    &  1088    &  0.91    &   3.24  \\ 
  GMC 78   &  22.8   &    7.8   &  3.34   &  1.11   &   4.64   &  1.92    &  1.88    &  1047    &  1.34    &   5.44  \\ 
  GMC 79   &  27.1   &   19.0   &  4.79   &  1.59   &   7.81   &  1.73    &  2.81    &  2236    &  1.18    &  12.93  \\ 
  GMC 80   &  16.7   &    2.0   &  1.98   &  0.66   &   2.17   &  2.06    &  1.05    &   345    &  1.72    &   1.67  \\ 
  GMC 81   &  45.0   &   23.0   &  4.09   &  1.36   &   6.21   &  4.25    &  2.35    &  1600    &   0.7    &   3.77  \\ 
  GMC 82   &  15.3   &    1.0   &  1.46   &  0.49   &    1.4   &  2.46    &  0.75    &   182    &  1.82    &   0.74  \\
  GMC 84   &  23.7   &    4.9   &   2.6   &  0.86   &   3.22   &  2.62    &  1.42    &   614    &  1.25    &   2.35  \\ 
  GMC 85   &  18.1   &    2.4   &  2.08   &  0.69   &   2.33   &  2.21    &  1.11    &   384    &   1.6    &   1.74  \\ 
  GMC 86   &  32.0   &   12.0   &   3.5   &  1.16   &   4.96   &  3.01    &  1.98    &  1153    &  0.96    &   3.83  \\ 
  GMC 87   &   7.5   &   0.35   &  1.24   &  0.41   &    1.1   &  1.03    &  0.62    &   127    &  3.63    &   1.24  \\ 
  GMC 88   &  46.0   &   20.0   &  3.77   &  1.25   &   5.52   &  4.75    &  2.15    &  1349    &  0.67    &   2.84  \\ 
  GMC 89   &  22.9   &    6.2   &  2.98   &  0.99   &   3.92   &  2.18    &  1.65    &   817    &  1.32    &   3.75  \\ 
  GMC 90   &  18.4   &    5.8   &  3.21   &  1.07   &   4.37   &  1.47    &   1.8    &   960    &  1.65    &   6.55  \\ 
  GMC 91   &  17.1   &    4.4   &   2.9   &  0.96   &   3.78   &  1.46    &  1.61    &   774    &  1.76    &   5.31  \\ 
  GMC 92   &  10.6   &   0.34   &  1.02   &  0.34   &   0.83   &  2.05    &   0.5    &    86    &  2.52    &   0.42  \\ 
  GMC 93   &  11.4   &    9.2   &  5.14   &  1.71   &   8.64   &  0.46    &  3.04    &  2595    &  2.82    &   57.0  \\ 
  GMC 94   &  16.0   &    5.6   &  3.38   &  1.12   &   4.72   &  1.13    &  1.91    &  1072    &  1.91    &   9.46  \\ 
  GMC 95   &  19.8   &    4.0   &  2.57   &  0.85   &   3.17   &  2.04    &   1.4    &   599    &   1.5    &   2.94  \\ 
  GMC 96   &   6.4   &   0.53   &  1.65   &  0.55   &   1.66   &  0.61    &  0.85    &   233    &   4.4    &   3.81  \\ 
  GMC 97   &  83.4   &  210.0   &  9.07   &  3.02   &  19.71   &  4.71    &  5.73    &  8648    &  0.41    &  18.36  \\ 
  GMC 98   &  28.2   &   12.0   &  3.73   &  1.24   &   5.44   &  2.35    &  2.13    &  1318    &   1.1    &    5.6  \\ 
  GMC 99   &   8.8   &   0.67   &  1.58   &  0.52   &   1.56   &  1.02    &  0.81    &   213    &  3.19    &    2.1  \\ 
 GMC 100   &  23.3   &   41.0   &  7.59   &  2.52   &   15.2   &  0.88    &  4.69    &  5919    &  1.44    &  67.62  \\ 
 GMC 101   &  11.6   &    1.4   &  1.99   &  0.66   &   2.18   &  1.21    &  1.05    &   348    &  2.48    &   2.88  \\ 
 GMC 102   &  29.8   &   42.0   &  6.79   &  2.26   &  12.95   &   1.4    &  4.15    &  4680    &  1.11    &  33.41  \\ 
 GMC 103   &   9.7   &   0.58   &   1.4   &  0.46   &   1.31   &  1.32    &  0.71    &   165    &  2.85    &   1.25  \\ 
 GMC 105   &   8.2   &    2.1   &  2.89   &  0.96   &   3.76   &   0.5    &   1.6    &   770    &  3.67    &  15.42  \\ 
 GMC 106   &  22.7   &    8.2   &  3.44   &  1.14   &   4.83   &  1.86    &  1.94    &  1109    &  1.35    &   5.96  \\ 
 GMC 107   &   5.7   &   0.46   &  1.62   &  0.54   &   1.63   &  0.52    &  0.84    &   227    &  4.94    &   4.34  \\ 
 GMC 109   &  25.4   &    5.3   &  2.61   &  0.87   &   3.24   &  2.88    &  1.43    &   620    &  1.17    &   2.15  \\ 
 GMC 110   &  19.3   &   12.0   &  4.51   &   1.5   &   7.15   &  1.12    &  2.63    &  1969    &  1.64    &   17.6  \\ 
 GMC 111   &   9.7   &    2.5   &   2.9   &  0.97   &   3.78   &  0.64    &  1.61    &   776    &   3.1    &  12.19  \\ 
 GMC 112   &  20.2   &    3.8   &  2.48   &  0.82   &   3.01   &  2.17    &  1.35    &   556    &  1.46    &   2.56  \\ 
 GMC 113   &   9.1   &   0.77   &  1.66   &  0.55   &   1.69   &  1.01    &  0.86    &   239    &   3.1    &   2.36  \\ 
 GMC 114   &  26.4   &   12.0   &  3.86   &  1.28   &    5.7   &  2.07    &  2.21    &  1414    &  1.18    &   6.84  \\ 
 GMC 115   &  12.0   &    4.2   &  3.38   &  1.12   &   4.72   &  0.75    &  1.91    &  1072    &  2.55    &  14.39  \\ 
 GMC 116   &  25.8   &    6.9   &  2.96   &  0.98   &   3.88   &  2.61    &  1.64    &   807    &  1.17    &    3.1  \\ 
 GMC 117   &  26.3   &   11.0   &   3.7   &  1.23   &   5.37   &  2.14    &  2.11    &  1295    &  1.18    &   6.04  \\ 
 GMC 118   &  23.3   &   12.0   &   4.1   &  1.36   &   6.24   &  1.62    &  2.36    &  1613    &  1.34    &   9.97  \\ 
 GMC 119   &  34.7   &    9.1   &  2.93   &  0.97   &   3.83   &  4.06    &  1.62    &   790    &  0.87    &   1.95  \\ 
 GMC 120   &   7.3   &   0.44   &   1.4   &  0.47   &   1.32   &  0.87    &  0.71    &   167    &  3.79    &   1.92  \\ 
 GMC 121   &  52.0   &   59.0   &  6.09   &  2.03   &  11.06   &  3.52    &  3.67    &  3721    &  0.63    &  10.57 \\ 
 GMC 122   &  53.7   &   55.0   &  5.79   &  1.92   &  10.27   &  3.88    &  3.47    &  3339    &  0.61    &    8.6  \\ 
 GMC 123   &  16.3   &    1.9   &  1.95   &  0.65   &   2.13   &  2.02    &  1.03    &   335    &  1.76    &   1.66  \\ 
 GMC 124   &  17.9   &    3.2   &  2.42   &   0.8   &    2.9   &  1.87    &  1.31    &   527    &  1.65    &   2.82  \\ 
 GMC 125   &  42.7   &   15.0   &  3.39   &  1.13   &   4.73   &  4.74    &  1.91    &  1076    &  0.72    &   2.27  \\ 
 GMC 126   &   7.2   &   0.36   &  1.28   &  0.43   &   1.15   &  0.93    &  0.64    &   137    &   3.8    &   1.46  \\ 
 GMC 127   &  34.0   &   12.0   &   3.4   &  1.13   &   4.75   &  3.39    &  1.92    &  1082    &   0.9    &   3.19  \\ 
 GMC 128   &  37.4   &   21.0   &  4.29   &  1.42   &   6.65   &  3.09    &  2.48    &  1768    &  0.84    &   5.72  \\ 
 GMC 129   &   9.6   &   0.41   &  1.18   &  0.39   &   1.03   &  1.54    &  0.59    &   116    &  2.82    &   0.75  \\ 
 GMC 130   &  12.8   &    1.0   &   1.6   &  0.53   &   1.59   &  1.73    &  0.83    &   219    &  2.19    &   1.27  \\ 
 GMC 131   &   8.4   &    0.7   &  1.65   &  0.55   &   1.67   &  0.91    &  0.86    &   235    &  3.36    &   2.59  \\ 
 GMC 133   &  16.1   &    4.5   &  3.02   &  1.01   &   4.01   &  1.28    &  1.68    &   845    &  1.88    &    6.6  \\ 
 GMC 134   &  25.8   &   15.0   &  4.36   &  1.45   &   6.81   &  1.77    &  2.53    &  1834    &  1.22    &  10.38  \\ 
 GMC 135   &  38.5   &    8.1   &  2.62   &  0.87   &   3.26   &  5.27    &  1.44    &   626    &  0.77    &   1.19  \\ 
 GMC 136   &  12.2   &   0.79   &  1.46   &  0.48   &   1.39   &  1.77    &  0.74    &   180    &  2.28    &   1.01  \\ 
 GMC 137   &  12.5   &    0.7   &  1.35   &  0.45   &   1.25   &  1.98    &  0.69    &   154    &   2.2    &   0.78  \\ 
 GMC 138   &  16.9   &    3.0   &  2.41   &   0.8   &   2.88   &  1.72    &  1.31    &   523    &  1.74    &   3.03  \\ 
 GMC 139   &   6.1   &   0.26   &  1.18   &  0.39   &   1.03   &  0.79    &  0.59    &   116    &  4.44    &   1.45  \\ 
 GMC 140   &  17.7   &    1.4   &  1.61   &  0.53   &   1.61   &  2.76    &  0.83    &   222    &  1.59    &    0.8  \\ 
 GMC 141   &  14.8   &    1.8   &  1.99   &  0.66   &   2.19   &  1.72    &  1.06    &   350    &  1.95    &   2.04  \\ 
 GMC 142   &  29.0   &   13.0   &  3.83   &  1.27   &   5.64   &  2.39    &  2.19    &  1393    &  1.07    &   5.84  \\ 
 GMC 143   &  10.5   &   0.64   &  1.41   &  0.47   &   1.33   &  1.47    &  0.72    &   169    &  2.64    &   1.15  \\ 
 GMC 144   &   9.2   &    1.7   &  2.46   &  0.82   &   2.97   &   0.7    &  1.33    &   545    &  3.21    &   7.84  \\ 
 GMC 145   &  17.0   &    1.9   &  1.91   &  0.64   &   2.06   &  2.19    &  1.01    &   320    &  1.69    &   1.46  \\ 
 GMC 146   &  23.3   &   16.0   &  4.74   &  1.58   &   7.69   &   1.4    &  2.78    &  2187    &  1.37    &  15.61  \\ 
 GMC 147   &  15.0   &    2.2   &  2.19   &  0.73   &   2.51   &  1.59    &  1.17    &   427    &  1.94    &   2.68  \\ 
 GMC 148   &  37.2   &   25.0   &  4.69   &  1.56   &   7.57   &   2.8    &  2.74    &  2138    &  0.86    &   7.62  \\ 
 GMC 149   &  19.5   &    8.3   &  3.73   &  1.24   &   5.44   &  1.37    &  2.13    &  1319    &  1.59    &   9.61  \\ 
 GMC 150   &  13.8   &    2.6   &  2.48   &  0.83   &   3.01   &  1.25    &  1.35    &   557    &  2.14    &   4.47  \\ 
 GMC 151   &  24.3   &   22.0   &  5.44   &  1.81   &   9.39   &   1.3    &  3.24    &  2930    &  1.33    &  22.58  \\ 
 GMC 152   &  26.5   &   22.0   &  5.21   &  1.73   &   8.82   &  1.54    &  3.09    &  2674    &  1.22    &  17.38  \\ 
 GMC 153   &  16.6   &    6.7   &  3.63   &  1.21   &   5.23   &  1.11    &  2.06    &  1247    &  1.86    &  11.19  \\ 
 GMC 154   &  16.3   &    3.2   &  2.53   &  0.84   &    3.1   &  1.56    &  1.38    &   582    &  1.82    &   3.74  \\ 
 GMC 155   &  31.6   &   18.0   &  4.32   &  1.43   &   6.71   &   2.4    &   2.5    &  1795    &   1.0    &   7.48  \\ 
 GMC 156   &   9.4   &   0.94   &  1.81   &   0.6   &    1.9   &  0.98    &  0.95    &   285    &  3.03    &   2.92  \\ 
 GMC 157   &  23.8   &    9.7   &  3.65   &  1.21   &   5.27   &  1.88    &  2.08    &  1260    &   1.3    &   6.71  \\ 
 GMC 158   &  41.1   &   37.0   &  5.43   &   1.8   &   9.36   &   2.8    &  3.23    &  2913    &  0.79    &  10.39  \\ 
 GMC 159   &  14.0   &    2.8   &  2.56   &  0.85   &   3.15   &  1.23    &   1.4    &   593    &  2.12    &   4.81  \\ 
 GMC 160   &  19.9   &    5.9   &  3.11   &  1.04   &   4.18   &  1.69    &  1.74    &   900    &  1.52    &   5.31  \\ 
 GMC 161   &  40.8   &   27.0   &  4.65   &  1.55   &   7.49   &  3.23    &  2.72    &  2103    &  0.78    &    6.5  \\ 
 GMC 162   &  34.9   &   83.0   &  8.82   &  2.93   &  18.91   &  1.36    &  5.55    &  8141    &  0.98    &  59.94  \\ 
 GMC 163   &  12.9   &    3.4   &  2.94   &  0.98   &   3.84   &  0.95    &  1.63    &   794    &  2.34    &   8.32  \\ 
 GMC 164   &  13.2   &    2.0   &  2.23   &  0.74   &   2.57   &   1.3    &  1.19    &   442    &  2.21    &    3.4  \\ 
 GMC 165   &   5.0   &   0.17   &  1.05   &  0.35   &   0.87   &  0.67    &  0.52    &    91    &  5.35    &   1.37  \\ 
 GMC 167   &  26.1   &    8.7   &   3.3   &   1.1   &   4.55   &  2.37    &  1.86    &  1018    &  1.17    &   4.29  \\ 
 GMC 168   &  44.7   &   27.0   &  4.44   &  1.48   &   7.01   &  3.87    &  2.58    &  1910    &  0.71    &   4.94  \\ 
 GMC 169   &  14.4   &    3.4   &  2.78   &  0.92   &   3.55   &  1.18    &  1.53    &   707    &  2.08    &   5.97  \\ 
 GMC 170   &  19.4   &    3.2   &  2.32   &  0.77   &   2.74   &  2.19    &  1.25    &   484    &  1.51    &   2.21  \\ 
 GMC 171   &  41.8   &  110.0   &  9.28   &  3.08   &  20.35   &  1.68    &  5.88    &  9061    &  0.82    &  53.93  \\ 
 GMC 172   &   8.8   &   0.69   &   1.6   &  0.53   &    1.6   &   1.0    &  0.83    &   220    &  3.19    &    2.2  \\ 
 GMC 173   &   8.4   &   0.26   &  1.01   &  0.33   &   0.81   &  1.49    &  0.49    &    82    &  3.17    &   0.55  \\ 
 GMC 174   &  36.2   &   22.0   &  4.46   &  1.48   &   7.04   &  2.83    &  2.59    &  1922    &  0.87    &   6.78  \\ 
 GMC 175   &  12.5   &    4.0   &  3.24   &  1.08   &   4.42   &  0.83    &  1.81    &   975    &  2.44    &  11.79  \\ 
 GMC 177   &  14.7   &    4.8   &  3.27   &  1.09   &   4.49   &  1.04    &  1.83    &   996    &  2.08    &    9.6  \\ 
 GMC 178   &  58.6   &   51.0   &  5.34   &  1.77   &   9.13   &  4.78    &  3.17    &  2811    &  0.55    &   5.87  \\ 
 GMC 179   &  13.6   &    1.6   &  1.96   &  0.65   &   2.14   &  1.54    &  1.04    &   338    &  2.11    &   2.19  \\ 
 GMC 180   &  21.9   &    8.9   &  3.65   &  1.21   &   5.26   &  1.66    &  2.07    &  1256    &  1.41    &   7.55  \\ 
 GMC 181   &  19.0   &    3.1   &  2.31   &  0.77   &   2.71   &  2.13    &  1.25    &   478    &  1.54    &   2.24  \\ 
 GMC 182   &  37.7   &   16.0   &  3.73   &  1.24   &   5.43   &   3.6    &  2.12    &  1315    &  0.82    &   3.65  \\ 
 GMC 183   &  15.0   &    4.0   &  2.95   &  0.98   &   3.87   &  1.18    &  1.64    &   804    &  2.01    &    6.8  \\ 
 GMC 184   &  44.4   &   14.0   &  3.21   &  1.07   &   4.37   &   5.3    &   1.8    &   960    &  0.69    &   1.81  \\ 
 GMC 185   &  23.0   &    2.8   &   2.0   &  0.66   &   2.19   &  3.27    &  1.06    &   351    &  1.25    &   1.07  \\ 
 GMC 186   &  34.0   &   11.0   &  3.25   &  1.08   &   4.46   &  3.54    &  1.82    &   987    &   0.9    &   2.78  \\ 
 GMC 187   &  35.7   &   11.0   &  3.17   &  1.06   &    4.3   &   3.9    &  1.78    &   937    &  0.85    &    2.4  \\ 
 GMC 188   &  32.9   &   12.0   &  3.45   &  1.15   &   4.86   &  3.18    &  1.95    &  1120    &  0.93    &   3.52  \\ 
 GMC 189   &  21.3   &    3.1   &  2.18   &  0.73   &    2.5   &  2.67    &  1.17    &   424    &  1.37    &   1.59  \\ 
 GMC 190   &  19.9   &    2.5   &  2.03   &  0.67   &   2.25   &   2.6    &  1.08    &   363    &  1.45    &   1.39  \\ 
 GMC 191   &  30.1   &    7.8   &  2.91   &  0.97   &   3.79   &  3.32    &  1.61    &   780    &   1.0    &   2.35  \\ 
 GMC 192   &  21.5   &    3.6   &  2.34   &  0.78   &   2.77   &  2.52    &  1.26    &   491    &  1.37    &   1.95  \\ 
 GMC 193   &  18.1   &   11.0   &  4.46   &  1.48   &   7.04   &  1.03    &  2.59    &  1922    &  1.75    &  18.66  \\ 
 GMC 195   &   7.6   &   0.79   &  1.84   &  0.61   &   1.96   &   0.7    &  0.97    &   297    &  3.76    &   4.23  \\ 
 GMC 196   &  16.1   &    3.4   &  2.63   &  0.87   &   3.27   &  1.47    &  1.44    &   628    &  1.85    &   4.26  \\ 
 GMC 198   &  15.7   &    3.6   &  2.74   &  0.91   &   3.47   &  1.36    &  1.51    &   685    &   1.9    &   5.03  \\ 
 GMC 202   &  36.0   &   14.0   &  3.57   &  1.19   &   5.09   &  3.51    &  2.02    &  1199    &  0.86    &   3.41  \\ 
 GMC 203   &  12.8   &   0.78   &  1.41   &  0.47   &   1.33   &  1.96    &  0.72    &   169    &  2.16    &   0.86  \\ 
 GMC 204   &   6.4   &   0.45   &  1.52   &   0.5   &   1.47   &  0.66    &  0.78    &   196    &  4.36    &   2.96  \\ 
 GMC 205   &  31.7   &    7.4   &  2.76   &  0.92   &   3.52   &  3.77    &  1.52    &   698    &  0.94    &   1.85  \\ 
 GMC 206   &  17.2   &    5.1   &  3.11   &  1.04   &   4.18   &  1.37    &  1.74    &   900    &  1.76    &   6.57  \\ 
 GMC 207   &  25.3   &    6.0   &  2.78   &  0.93   &   3.56   &  2.69    &  1.53    &   710    &  1.18    &   2.64  \\ 
 GMC 208   &  13.9   &    3.7   &  2.95   &  0.98   &   3.87   &  1.06    &  1.64    &   803    &  2.17    &   7.58  \\ 
 GMC 209   &  31.8   &   12.0   &  3.51   &  1.17   &   4.98   &  2.98    &  1.99    &  1161    &  0.97    &    3.9  \\ 
 GMC 210   &   6.4   &    0.3   &  1.24   &  0.41   &    1.1   &  0.81    &  0.62    &   128    &  4.26    &   1.57  \\ 
 GMC 211   &  20.8   &    7.7   &  3.48   &  1.16   &   4.91   &  1.62    &  1.97    &  1138    &  1.48    &   7.04  \\ 
 GMC 212   &  10.8   &    3.2   &  3.11   &  1.03   &   4.18   &  0.69    &  1.74    &   899    &  2.81    &  12.95  \\ 
 GMC 213   &  10.8   &    2.3   &  2.64   &  0.88   &   3.29   &  0.82    &  1.44    &   634    &  2.76    &   7.74  \\ 
 GMC 214   &  90.8   &  120.0   &  6.57   &  2.19   &  12.36   &  7.36    &   4.0    &  4372    &  0.36    &   5.94  \\ 
 GMC 215   &  12.5   &   0.84   &  1.48   &  0.49   &   1.43   &   1.8    &  0.76    &   187    &  2.23    &   1.04  \\ 
 GMC 216   &  11.4   &    1.3   &  1.93   &  0.64   &   2.09   &  1.21    &  1.02    &   327    &  2.52    &    2.7  \\ 
 GMC 217   &  42.6   &   25.0   &  4.38   &  1.46   &   6.86   &  3.66    &  2.54    &  1852    &  0.74    &   5.06  \\ 
 GMC 218   &  25.3   &    6.7   &  2.94   &  0.98   &   3.85   &  2.54    &  1.63    &   798    &  1.19    &   3.14  \\ 
 GMC 219   &  20.6   &    6.3   &  3.16   &  1.05   &   4.28   &  1.75    &  1.77    &   930    &  1.48    &    5.3  \\ 
 GMC 220   &  15.9   &    5.5   &  3.36   &  1.12   &   4.68   &  1.13    &  1.89    &  1059    &  1.93    &   9.37  \\ 
 GMC 221   &  16.9   &    5.7   &  3.32   &   1.1   &   4.59   &  1.25    &  1.87    &  1031    &  1.81    &   8.24  \\ 
 GMC 223   &   7.1   &   0.68   &  1.77   &  0.59   &   1.84   &  0.66    &  0.93    &   272    &   4.0    &   4.11  \\ 
 GMC 224   &  46.1   &   46.0   &  5.71   &   1.9   &  10.08   &  3.15    &  3.42    &  3248    &  0.71    &  10.32  \\ 
 GMC 225   &  47.0   &   44.0   &  5.53   &  1.84   &   9.62   &  3.34    &   3.3    &  3036    &  0.69    &   9.08  \\ 
 GMC 226   &  13.6   &    5.8   &  3.73   &  1.24   &   5.44   &  0.81    &  2.13    &  1321    &  2.28    &  16.31  \\ 
 GMC 227   &  19.8   &    6.8   &  3.35   &  1.11   &   4.65   &  1.56    &  1.89    &  1051    &  1.55    &   6.73  \\ 
 GMC 229   &  17.7   &    2.0   &  1.92   &  0.64   &   2.08   &  2.31    &  1.01    &   324    &  1.62    &    1.4  \\ 
 GMC 232   &  32.0   &   13.0   &  3.65   &  1.21   &   5.26   &   2.9    &  2.07    &  1255    &  0.97    &   4.34  \\ 
 GMC 234   &  11.9   &    1.9   &  2.29   &  0.76   &   2.67   &  1.09    &  1.23    &   467    &  2.46    &   4.29  \\ 
 GMC 236   &   7.3   &   0.77   &  1.86   &  0.62   &   1.98   &  0.66    &  0.98    &   301    &  3.91    &   4.59  \\ 
 GMC 237   &  31.3   &   10.0   &  3.23   &  1.07   &   4.42   &  3.16    &  1.81    &   974    &  0.97    &   3.08  \\ 
 GMC 238   &  12.3   &    1.9   &  2.25   &  0.75   &   2.61   &  1.16    &  1.21    &   451    &  2.38    &   3.88  \\ 
 GMC 240   &   9.0   &    1.3   &  2.17   &  0.72   &   2.48   &  0.76    &  1.16    &   420    &  3.23    &   5.52  \\ 
 GMC 241   &  15.9   &    2.3   &  2.18   &  0.72   &   2.49   &  1.75    &  1.16    &   421    &  1.83    &   2.41  \\ 
 GMC 242   &  12.8   &    5.0   &  3.57   &  1.19   &   5.11   &  0.77    &  2.03    &  1204    &  2.41    &  15.54  \\ 
 GMC 243   &  28.2   &   11.0   &  3.57   &  1.19   &    5.1   &  2.46    &  2.03    &  1202    &  1.09    &   4.89  \\ 
\enddata

\end{deluxetable*}

\bibliography{APJ}{}

\begin{thebibliography}{}
\expandafter\ifx\csname natexlab\endcsname\relax\def\natexlab#1{#1}\fi
\providecommand{\url}[1]{\href{#1}{#1}}
\providecommand{\dodoi}[1]{doi:~\href{http://doi.org/#1}{\nolinkurl{#1}}}
\providecommand{\doeprint}[1]{\href{http://ascl.net/#1}{\nolinkurl{http://ascl.net/#1}}}
\providecommand{\doarXiv}[1]{\href{https://arxiv.org/abs/#1}{\nolinkurl{https://arxiv.org/abs/#1}}}

\bibitem[{{Ballesteros-Paredes} {et~al.}(2020){Ballesteros-Paredes}, {Andr{\'e}}, {Hennebelle}, {Klessen}, {Kruijssen}, {Chevance}, {Nakamura}, {Adamo}, \& {V{\'a}zquez-Semadeni}}]{Bal20}
{Ballesteros-Paredes}, J., {Andr{\'e}}, P., {Hennebelle}, P., {et~al.} 2020, \ssr, 216, 76, \dodoi{10.1007/s11214-020-00698-3}

\bibitem[{{Bertoldi} \& {McKee}(1992)}]{Ber92}
{Bertoldi}, F., \& {McKee}, C.~F. 1992, \apj, 395, 140, \dodoi{10.1086/171638}

\bibitem[{{Chevance} {et~al.}(2023){Chevance}, {Krumholz}, {McLeod}, {Ostriker}, {Rosolowsky}, \& {Sternberg}}]{Che23}
{Chevance}, M., {Krumholz}, M.~R., {McLeod}, A.~F., {et~al.} 2023, in Astronomical Society of the Pacific Conference Series, Vol. 534, Protostars and Planets VII, ed. S.~{Inutsuka}, Y.~{Aikawa}, T.~{Muto}, K.~{Tomida}, \& M.~{Tamura}, 1, \dodoi{10.48550/arXiv.2203.09570}

\bibitem[{{Chevance} {et~al.}(2020){Chevance}, {Kruijssen}, {Hygate}, {Schruba}, {Longmore}, {Groves}, {Henshaw}, {Herrera}, {Hughes}, {Jeffreson}, {Lang}, {Leroy}, {Meidt}, {Pety}, {Razza}, {Rosolowsky}, {Schinnerer}, {Bigiel}, {Blanc}, {Emsellem}, {Faesi}, {Glover}, {Haydon}, {Ho}, {Kreckel}, {Lee}, {Liu}, {Querejeta}, {Saito}, {Sun}, {Usero}, \& {Utomo}}]{Che20}
{Chevance}, M., {Kruijssen}, J.~M.~D., {Hygate}, A. P.~S., {et~al.} 2020, \mnras, 493, 2872, \dodoi{10.1093/mnras/stz3525}

\bibitem[{{Chomiuk} \& {Povich}(2011)}]{Cho11}
{Chomiuk}, L., \& {Povich}, M.~S. 2011, \aj, 142, 197, \dodoi{10.1088/0004-6256/142/6/197}

\bibitem[{{Davies} {et~al.}(2011){Davies}, {Hoare}, {Lumsden}, {Hosokawa}, {Oudmaijer}, {Urquhart}, {Mottram}, \& {Stead}}]{Dav11}
{Davies}, B., {Hoare}, M.~G., {Lumsden}, S.~L., {et~al.} 2011, \mnras, 416, 972, \dodoi{10.1111/j.1365-2966.2011.19095.x}

\bibitem[{{Elia} {et~al.}(2022){Elia}, {Molinari}, {Schisano}, {Soler}, {Merello}, {Russeil}, {Veneziani}, {Zavagno}, {Noriega-Crespo}, {Olmi}, {Benedettini}, {Hennebelle}, {Klessen}, {Leurini}, {Paladini}, {Pezzuto}, {Traficante}, {Eden}, {Martin}, {Sormani}, {Coletta}, {Colman}, {Plume}, {Maruccia}, {Mininni}, \& {Liu}}]{Eli22}
{Elia}, D., {Molinari}, S., {Schisano}, E., {et~al.} 2022, \apj, 941, 162, \dodoi{10.3847/1538-4357/aca27d}

\bibitem[{{Ferri{\`e}re}(2001)}]{Fer01}
{Ferri{\`e}re}, K.~M. 2001, Reviews of Modern Physics, 73, 1031, \dodoi{10.1103/RevModPhys.73.1031}

\bibitem[{{Friesen} {et~al.}(2017){Friesen}, {Pineda}, {co-PIs}, {Rosolowsky}, {Alves}, {Chac{\'o}n-Tanarro}, {How-Huan Chen}, {Chun-Yuan Chen}, {Di Francesco}, {Keown}, {Kirk}, {Punanova}, {Seo}, {Shirley}, {Ginsburg}, {Hall}, {Offner}, {Singh}, {Arce}, {Caselli}, {Goodman}, {Martin}, {Matzner}, {Myers}, {Redaelli}, \& {GAS Collaboration}}]{Fri17}
{Friesen}, R.~K., {Pineda}, J.~E., {co-PIs}, {et~al.} 2017, \apj, 843, 63, \dodoi{10.3847/1538-4357/aa6d58}

\bibitem[{{Goodman} {et~al.}(1998){Goodman}, {Barranco}, {Wilner}, \& {Heyer}}]{Goo98}
{Goodman}, A.~A., {Barranco}, J.~A., {Wilner}, D.~J., \& {Heyer}, M.~H. 1998, \apj, 504, 223, \dodoi{10.1086/306045}

\bibitem[{{Grudi{\'c}} {et~al.}(2022){Grudi{\'c}}, {Guszejnov}, {Offner}, {Rosen}, {Raju}, {Faucher-Gigu{\`e}re}, \& {Hopkins}}]{Gru22}
{Grudi{\'c}}, M.~Y., {Guszejnov}, D., {Offner}, S. S.~R., {et~al.} 2022, \mnras, 512, 216, \dodoi{10.1093/mnras/stac526}

\bibitem[{{Heyer} \& {Dame}(2015)}]{Hey15}
{Heyer}, M., \& {Dame}, T.~M. 2015, \araa, 53, 583, \dodoi{10.1146/annurev-astro-082214-122324}

\bibitem[{{Heyer} {et~al.}(2009){Heyer}, {Krawczyk}, {Duval}, \& {Jackson}}]{Hey09}
{Heyer}, M., {Krawczyk}, C., {Duval}, J., \& {Jackson}, J.~M. 2009, \apj, 699, 1092, \dodoi{10.1088/0004-637X/699/2/1092}

\bibitem[{{Jeffreson} {et~al.}(2021){Jeffreson}, {Keller}, {Winter}, {Chevance}, {Kruijssen}, {Krumholz}, \& {Fujimoto}}]{Jef21}
{Jeffreson}, S. M.~R., {Keller}, B.~W., {Winter}, A.~J., {et~al.} 2021, \mnras, 505, 1678, \dodoi{10.1093/mnras/stab1293}

\bibitem[{{Kainulainen} {et~al.}(2013){Kainulainen}, {Ragan}, {Henning}, \& {Stutz}}]{Kai13}
{Kainulainen}, J., {Ragan}, S., {Henning}, T., \& {Stutz}, A. 2013, in Protostars and Planets VI Posters

\bibitem[{{Kainulainen} {et~al.}(2017){Kainulainen}, {Stutz}, {Stanke}, {Abreu-Vicente}, {Beuther}, {Henning}, {Johnston}, \& {Megeath}}]{Kai17}
{Kainulainen}, J., {Stutz}, A.~M., {Stanke}, T., {et~al.} 2017, \aap, 600, A141, \dodoi{10.1051/0004-6361/201628481}

\bibitem[{{Kennicutt}(1998)}]{Ken98}
{Kennicutt}, Robert~C., J. 1998, \apj, 498, 541, \dodoi{10.1086/305588}

\bibitem[{{Kennicutt} \& {Evans}(2012)}]{Ken12}
{Kennicutt}, R.~C., \& {Evans}, N.~J. 2012, \araa, 50, 531, \dodoi{10.1146/annurev-astro-081811-125610}

\bibitem[{{Klessen} \& {Hennebelle}(2010)}]{Kle10}
{Klessen}, R.~S., \& {Hennebelle}, P. 2010, \aap, 520, A17, \dodoi{10.1051/0004-6361/200913780}

\bibitem[{{Krumholz} \& {McKee}(2005)}]{Kru05}
{Krumholz}, M.~R., \& {McKee}, C.~F. 2005, \apj, 630, 250, \dodoi{10.1086/431734}

\bibitem[{{Krumholz} \& {Tan}(2007)}]{Kru07}
{Krumholz}, M.~R., \& {Tan}, J.~C. 2007, \apj, 654, 304, \dodoi{10.1086/509101}

\bibitem[{{Lada} {et~al.}(2010){Lada}, {Lombardi}, \& {Alves}}]{Lad10}
{Lada}, C.~J., {Lombardi}, M., \& {Alves}, J.~F. 2010, \apj, 724, 687, \dodoi{10.1088/0004-637X/724/1/687}

\bibitem[{{Larson}(1981)}]{Lar81}
{Larson}, R.~B. 1981, \mnras, 194, 809, \dodoi{10.1093/mnras/194.4.809}

\bibitem[{{Li}(2018)}]{Li18}
{Li}, G.-X. 2018, \mnras, 477, 4951, \dodoi{10.1093/mnras/sty657}

\bibitem[{{Li} {et~al.}(2024){Li}, {Sanhueza}, {Beuther}, {Chen}, {Kuiper}, {Olguin}, {Pudritz}, {Stephens}, {Zhang}, {Nakamura}, {Lu}, {Kuruwita}, {Sakai}, {Henning}, {Taniguchi}, \& {Li}}]{Li24}
{Li}, S., {Sanhueza}, P., {Beuther}, H., {et~al.} 2024, Nature Astronomy, 8, 472, \dodoi{10.1038/s41550-023-02181-9}

\bibitem[{{Licquia} \& {Newman}(2015)}]{Lic15}
{Licquia}, T.~C., \& {Newman}, J.~A. 2015, \apj, 806, 96, \dodoi{10.1088/0004-637X/806/1/96}

\bibitem[{{Lu} {et~al.}(2022){Lu}, {Li}, {Zhang}, \& {Lin}}]{Lu22}
{Lu}, X., {Li}, G.-X., {Zhang}, Q., \& {Lin}, Y. 2022, Nature Astronomy, 6, 837, \dodoi{10.1038/s41550-022-01681-4}

\bibitem[{{Luo} {et~al.}(2024){Luo}, {Liu}, {Li}, {Pan}, \& {Yang}}]{Luo24}
{Luo}, A.-X., {Liu}, H.-L., {Li}, G.-X., {Pan}, S., \& {Yang}, D.-T. 2024, Research in Astronomy and Astrophysics, 24, 065003, \dodoi{10.1088/1674-4527/ad3ec8}

\bibitem[{{Matzner} \& {McKee}(2000)}]{Maz00}
{Matzner}, C.~D., \& {McKee}, C.~F. 2000, \apj, 545, 364, \dodoi{10.1086/317785}

\bibitem[{{McKee} \& {Ostriker}(2007)}]{Mck07}
{McKee}, C.~F., \& {Ostriker}, E.~C. 2007, \araa, 45, 565, \dodoi{10.1146/annurev.astro.45.051806.110602}

\bibitem[{{McKee} \& {Tan}(2003)}]{Mck03}
{McKee}, C.~F., \& {Tan}, J.~C. 2003, \apj, 585, 850, \dodoi{10.1086/346149}

\bibitem[{{Padoan} \& {Nordlund}(2002)}]{Pad02}
{Padoan}, P., \& {Nordlund}, {\r{A}}. 2002, \apj, 576, 870, \dodoi{10.1086/341790}

\bibitem[{{Padoan} \& {Nordlund}(2011)}]{Pad11}
---. 2011, \apj, 730, 40, \dodoi{10.1088/0004-637X/730/1/40}

\bibitem[{{Padoan} {et~al.}(2020){Padoan}, {Pan}, {Juvela}, {Haugb{\o}lle}, \& {Nordlund}}]{Pad20}
{Padoan}, P., {Pan}, L., {Juvela}, M., {Haugb{\o}lle}, T., \& {Nordlund}, {\r{A}}. 2020, \apj, 900, 82, \dodoi{10.3847/1538-4357/abaa47}

\bibitem[{{Peretto} {et~al.}(2023){Peretto}, {Rigby}, {Louvet}, {Fuller}, {Traficante}, \& {Gaudel}}]{Per23}
{Peretto}, N., {Rigby}, A.~J., {Louvet}, F., {et~al.} 2023, \mnras, 525, 2935, \dodoi{10.1093/mnras/stad2453}

\bibitem[{{Pineda} {et~al.}(2010){Pineda}, {Goodman}, {Arce}, {Caselli}, {Foster}, {Myers}, \& {Rosolowsky}}]{Pin10}
{Pineda}, J.~E., {Goodman}, A.~A., {Arce}, H.~G., {et~al.} 2010, \apjl, 712, L116, \dodoi{10.1088/2041-8205/712/1/L116}

\bibitem[{{Shu}(1977)}]{Shu77}
{Shu}, F.~H. 1977, \apj, 214, 488, \dodoi{10.1086/155274}

\bibitem[{{Shu} {et~al.}(1987){Shu}, {Adams}, \& {Lizano}}]{Shu87}
{Shu}, F.~H., {Adams}, F.~C., \& {Lizano}, S. 1987, \araa, 25, 23, \dodoi{10.1146/annurev.aa.25.090187.000323}

\bibitem[{{Staff} {et~al.}(2019){Staff}, {Tanaka}, \& {Tan}}]{Sta19}
{Staff}, J.~E., {Tanaka}, K. E.~I., \& {Tan}, J.~C. 2019, \apj, 882, 123, \dodoi{10.3847/1538-4357/ab36b3}

\bibitem[{{Tanaka} {et~al.}(2017){Tanaka}, {Tan}, \& {Zhang}}]{Tan17}
{Tanaka}, K. E.~I., {Tan}, J.~C., \& {Zhang}, Y. 2017, \apj, 835, 32, \dodoi{10.3847/1538-4357/835/1/32}

\bibitem[{{V{\'a}zquez-Semadeni} {et~al.}(2024{\natexlab{a}}){V{\'a}zquez-Semadeni}, {G{\'o}mez}, \& {Gonz{\'a}lez-Samaniego}}]{Vaz24a}
{V{\'a}zquez-Semadeni}, E., {G{\'o}mez}, G.~C., \& {Gonz{\'a}lez-Samaniego}, A. 2024{\natexlab{a}}, \mnras, 530, 3445, \dodoi{10.1093/mnras/stae1090}

\bibitem[{{V{\'a}zquez-Semadeni} {et~al.}(2019){V{\'a}zquez-Semadeni}, {Palau}, {Ballesteros-Paredes}, {G{\'o}mez}, \& {Zamora-Avil{\'e}s}}]{Vaz19}
{V{\'a}zquez-Semadeni}, E., {Palau}, A., {Ballesteros-Paredes}, J., {G{\'o}mez}, G.~C., \& {Zamora-Avil{\'e}s}, M. 2019, \mnras, 490, 3061, \dodoi{10.1093/mnras/stz2736}

\bibitem[{{V{\'a}zquez-Semadeni} {et~al.}(2024{\natexlab{b}}){V{\'a}zquez-Semadeni}, {Palau}, {G{\'o}mez}, {Arroyo-Ch{\'a}vez}, {Alig}, {Ballesteros-Paredes}, {Camacho}, {Gonz{\'a}lez-Samaniego}, \& {Burkert}}]{Vaz24b}
{V{\'a}zquez-Semadeni}, E., {Palau}, A., {G{\'o}mez}, G.~C., {et~al.} 2024{\natexlab{b}}, arXiv e-prints, arXiv:2408.10406, \dodoi{10.48550/arXiv.2408.10406}

\bibitem[{{Xu} {et~al.}(2023){Xu}, {Wang}, {He}, {Wu}, {Zhu}, \& {Mardones}}]{Xu23}
{Xu}, F., {Wang}, K., {He}, Y., {et~al.} 2023, \apjs, 269, 38, \dodoi{10.3847/1538-4365/acfee2}

\bibitem[{{Yang} {et~al.}(2023){Yang}, {Chen}, {Jiang}, {Chen}, {Yu}, \& {Li}}]{Yan23}
{Yang}, Y., {Chen}, X., {Jiang}, Z., {et~al.} 2023, \apj, 955, 154, \dodoi{10.3847/1538-4357/aced09}

\bibitem[{{Zamora-Avil{\'e}s} \& {V{\'a}zquez-Semadeni}(2014)}]{Zam14}
{Zamora-Avil{\'e}s}, M., \& {V{\'a}zquez-Semadeni}, E. 2014, \apj, 793, 84, \dodoi{10.1088/0004-637X/793/2/84}

\bibitem[{{Zhang} {et~al.}(2024){Zhang}, {Zhou}, {Esimbek}, {Baan}, {Tang}, {Li}, {He}, {Wu}, {Zhou}, {Ma}, {Tursun}, {Ji}, {Chang}, {Li}, \& {Komesh}}]{Zha24}
{Zhang}, W., {Zhou}, J., {Esimbek}, J., {et~al.} 2024, \apjs, 275, 7, \dodoi{10.3847/1538-4365/ad7828}

\end{thebibliography}
\bibliographystyle{aasjournal}

\end{document}